\documentclass[aps,prc,reprint,amsmath]{revtex4-1}

\usepackage{hyperref}
\usepackage{graphicx}
\graphicspath{{fig/}}

\usepackage{tikz}
\usetikzlibrary{shapes,calc,matrix,external}
\newcommand{\psec}[1]{\phantomsection\addcontentsline{toc}{section}{#1}}
\newcommand{\trento}{T\raisebox{-.5ex}{R}ENTo}
\newcommand{\avg}[1]{\langle #1 \rangle}
\newcommand{\nch}{N_\text{ch}}
\newcommand{\eccratio}{\sqrt{\langle \varepsilon_2^2 \rangle}/\sqrt{\langle \varepsilon_3^2 \rangle}^{\,0.6}}

\begin{document}

\title{Alternative ansatz to wounded nucleon and binary collision scaling \\ in high-energy nuclear collisions}

\author{J.\ Scott Moreland}
\author{Jonah E.\ Bernhard}
\author{Steffen A.\ Bass}
\affiliation{Department of Physics, Duke University, Durham, NC 27708-0305}

\date{\today}

\begin{abstract}
  We introduce \trento, a new parametric initial condition model for high-energy nuclear collisions based on eikonal entropy deposition via a ``reduced thickness'' function.
  The model simultaneously describes experimental proton-proton, proton-nucleus, and nucleus-nucleus multiplicity distributions, and generates nucleus-nucleus eccentricity harmonics consistent with experimental flow constraints.
  In addition, the model is compatible with ultra-central uranium-uranium data unlike existing models that include binary collision terms.
\end{abstract}

\maketitle

\psec{Introduction}

Over the last decade, the ultra-relativistic heavy-ion collision programs at the Relativistic Heavy Ion Collider (RHIC) and the Large Hadron Collider (LHC) have succeeded in producing and exploring a novel, highly excited phase of QCD matter dubbed the strongly-interacting Quark-Gluon Plasma (sQGP)
\cite{Arsene:2004fa,Adcox:2004mh,Back:2004je,Adams:2005dq,Gyulassy:2004zy,Muller:2006ee,Muller:2012zq}.
A major goal of current research is the quantification of fundamental sQGP properties, typically accomplished by matching experimental measurements to computational models for the full spacetime evolution of heavy-ion collisions \cite{Petersen:2010zt,Novak:2013bqa}.
While viscous relativistic fluid dynamics provides a stable, well-tested description of the thermalized sQGP medium \cite{Baier:2006gy,Song:2007ux,Luzum:2008cw,Schenke:2010rr,Shen:2011eg,Shen:2014vra}, the initial state of the collision remains poorly constrained and constitutes the largest source of uncertainty in modern computational models \cite{Song:2010mg, Retinskaya:2013gca}.

Initial condition models generate profiles of energy or entropy at the sQGP thermalization time to be evolved by fluid dynamics.
This is accomplished by two general approaches:
dynamical models, which explicitly simulate the initial state and pre-equilibrium evolution of the collision \cite{Schenke:2012wb,vanderSchee:2013pia,Berges:2014yta,Kurkela:2014tea};
and simpler non-dynamical models, which neglect pre-equilibrium evolution and construct static profiles at the thermalization time.

The most successful dynamical model is IP-Glasma \cite{Schenke:2012wb}.
IP-Glasma uses weakly-coupled color-glass condensate (CGC) effective field theory \cite{McLerran:1993ni,McLerran:1993ka,Gelis:2010nm} and classical Yang-Mills evolution to quantitatively describe the latest event-by-event data on higher-order flow harmonics and a variety of other observables \cite{Schenke:2014zha}.
More recently, strongly-coupled AdS/CFT holography was applied to create a dynamical simulation of central nuclear collisions at the LHC \cite{vanderSchee:2013pia}.
While these models are all based on approximations of QCD or related quantum field theories, they encounter limits in their application to the full range of collision systems \cite{Schenke:2013dpa,vanderSchee:2013pia}, and their explicit treatment of pre-equilibrium dynamics comes at a significant computational cost.

Non-dynamical models generate initial conditions directly at the thermalization time by asserting an ansatz for entropy deposition.
Although they cannot explain sQGP formation, they provide the necessary input for fluid dynamics and can constrain the outcome of ab-initio initial condition calculations.
The most widely-used prescription is the two-component Monte Carlo Glauber model, which determines participating nucleons via optical overlap and deposits energy or entropy for each participant and binary nucleon-nucleon collision.
Despite its simplicity, the Glauber model has qualitatively fit many experimental measurements \cite{Miller:2007ri} and inspired a number of similar models.
Notably, a participant quark model was proposed to describe the transverse-energy distributions of proton-proton, deuteron-gold, and gold-gold collisions at RHIC without invoking binary collision scaling \cite{PhysRevC.89.044905}.
This may suggest that the two-component wounded nucleon and binary collision ansatz is merely a proxy for participant quark scaling in nucleus-nucleus collisions.

In this work we introduce \trento, a new initial condition model for high-energy proton-proton, proton-nucleus, and nucleus-nucleus collisions.
It is an \emph{effective} model, intended to generate realistic Monte Carlo initial entropy profiles without assuming specific physical mechanisms for entropy production, pre-equilibrium dynamics, or thermalization.

\psec{Model}

Suppose a pair of projectiles labeled $A, B$ collide along beam axis $z$, and let $\rho^\text{part}_{A,B}$ be the density of nuclear matter that participates in inelastic collisions.
Each projectile may then be represented by its \emph{participant} thickness
\begin{equation}
  T_{A,B}(x, y) = \int dz \, \rho^\text{part}_{A,B}(x, y, z).
  \label{eq:thickness}
\end{equation}
The construction of these thickness functions will be addressed shortly; first, we postulate the following:
\begin{enumerate}
  \item The eikonal approximation is valid:  entropy is produced if $T_A$ and $T_B$ eikonally overlap.
  \item There exists a scalar field $f(T_A, T_B)$ which converts projectile thicknesses into entropy
    deposition.
\end{enumerate}
The function $f$ is proportional to the entropy created at mid-rapidity and at the hydrodynamic thermalization
time:
\begin{equation}
  f \propto dS/dy \, |_{\tau = \tau_0}.
\end{equation}
It should provide an effective description of early collision dynamics:
it need not arise from a first-principles calculation, but it must obey basic physical constraints.

Perhaps the simplest such function is a sum, $f~\sim~T_A~+~T_B$, in fact this is equivalent to a wounded nucleon model since the present thickness functions \eqref{eq:thickness} only include participant matter.
The two-component Glauber ansatz adds a quadratic term to account for binary collisions, i.e.\ $f \sim (T_A + T_B) + \alpha \, T_A T_B$.

However, recent results from ultra-central uranium-uranium collisions at RHIC \cite{FortheSTAR:2013bza,Wang:2014qxa} show that particle production does not scale with the number of binary collisions, excluding the two-component Glauber ansatz \cite{Goldschmidt:2015qya}.
Therefore $N$ \mbox{one-on-one} nucleon collisions should produce the same amount of entropy as a single \mbox{$N$-on-$N$} collision, which is mathematically equivalent to the function $f$ being scale-invariant:
\begin{equation}
  f(c \, T_A, c \, T_B) = c \, f(T_A, T_B)
  \label{eq:scale-inv}
\end{equation}
for any nonzero constant $c$.
Note, this is clearly broken by the binary collision term $(\alpha \, T_A T_B)$.
We will justify this constraint later in the text; for the moment we take it as a postulate.

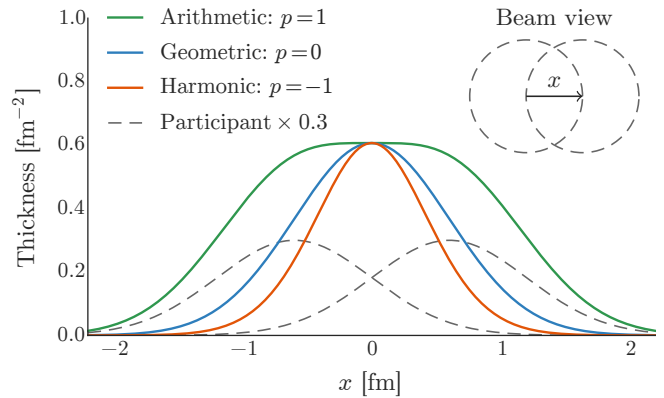
\begin{figure}[t]
  \definecolor{offblack}{HTML}{262626}
  \tikzsetnextfilename{reduced_thickness_final}
  \begin{tikzpicture}[
      scale=.5,
      color=offblack,
      greydash/.style={line width=0.6, dash pattern=on 5pt off 3pt, color=black!60!white}
    ]
    \def\r{1.5};
    \node (fig) {\includegraphics{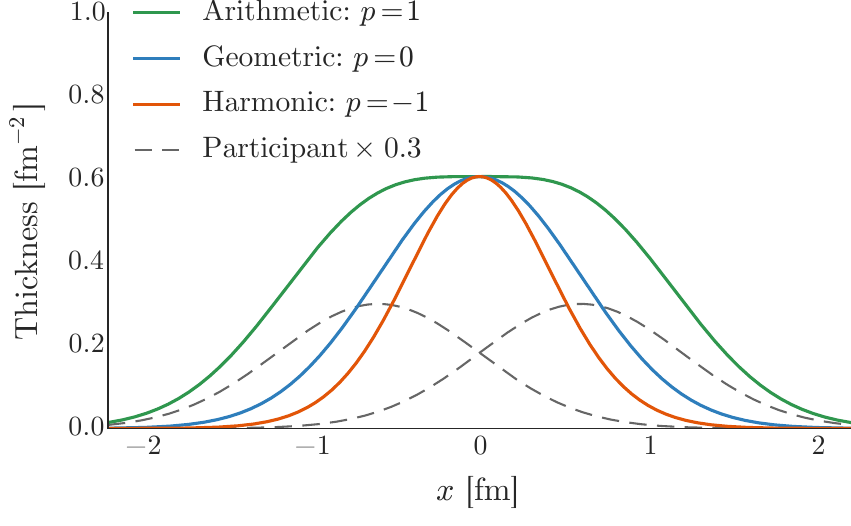}};
    \coordinate (a) at (5.2, 3.0);
    \coordinate (b) at ($(a) + (\r, 0)$);
    \node at ($(a)!0.5!(b) + (0, 1.4*\r)$) {\small Beam view};
    \draw[greydash] (a) circle (\r);
    \draw[greydash] (b) circle (\r);
    \draw[->, semithick] (a) -- node[above] {$\small x$} (b);
  \end{tikzpicture}
  \caption{
    \label{fig:TR}
    Reduced thickness of a pair of nucleon participants.
    The nucleons collide with a nonzero impact parameter along the $x$-direction as shown in the upper right.
    The grey dashed lines are one-dimensional cross sections of the participant nucleon thickness functions $T_A, T_B$, and the colored lines are the reduced thickness $T_R$ for $p = 1, 0, -1$ (green, blue, orange).
  }
\end{figure}

With these constraints in mind, we propose for $f$ the \emph{reduced thickness}
\begin{equation}
  f = T_R(p; T_A, T_B) \equiv \biggl( \frac{T_A^p + T_B^p}{2} \biggr)^{1/p},
  \label{eq:tr}
\end{equation}
so named because it takes two thicknesses $T_A, T_B$ and ``reduces'' them to a third thickness, similar to a
reduced mass.
This functional form---known as the generalized mean---interpolates between the minimum and maximum of $T_A, T_B$ depending on the value of the dimensionless parameter $p$, and simplifies to the arithmetic, geometric, and harmonic means for certain values:
\begin{equation}
  \newlength{\extraspace}
  \setlength{\extraspace}{0.5ex}
  T_R =
  \begin{cases}
    \max(T_A, T_B) & p \rightarrow +\infty, \\[\extraspace]
    (T_A + T_B)/2 & p = +1, \hfill \text{ (arithmetic)} \\[\extraspace]
    \sqrt{T_A T_B} & p = 0, \hfill \text{ (geometric)} \\[\extraspace]
    2\, T_A T_B/(T_A + T_B) & p = -1, \hfill \text{ (harmonic)} \\[\extraspace]
    \min(T_A, T_B) & p \rightarrow -\infty.
  \end{cases}
\end{equation}
Physically, $p$ interpolates among qualitatively different physical mechanisms for entropy production.
To see this, consider a pair of nucleon participants colliding with some nonzero impact parameter, as shown in Fig.~\ref{fig:TR}.
For $p = 1$, the reduced thickness is equivalent to a Monte Carlo wounded nucleon model and deposits a blob of entropy for each nucleon,
while for $p = 0$, the model deposits a single roughly symmetric blob at the midpoint of the collision,
and as $p$ becomes negative, it suppresses entropy deposition along the direction of the impact parameter.
Similar behavior was discussed in the context of small collision systems in \cite{Bzdak:2013zma}.
Note that the values $1, 0, -1$ are only special cases---$p$ is a continuous parameter---and the scale-invariant constraint \eqref{eq:scale-inv} is always satisfied.

We now detail the construction of the thickness functions $T_{A,B}(x, y)$, which combined with the definition
of the reduced thickness completes the specification of the model.  The procedure is constructed from the
ground up to handle a variety of collision systems; we begin with the simplest case.

Consider a collision of two protons $A, B$ with impact parameter $b$ along the $x$-direction and nuclear densities
\begin{equation}
  \rho_{A,B} = \rho_\text{proton}(x \pm b/2, y, z),
\end{equation}
and assume that the integral $\int dz \, \rho_\text{proton}$ either has a closed form or may be evaluated numerically, so that the proton thickness functions can be calculated.
The protons collide with probability \cite{dEnterria:2010hd}
\begin{equation}
  P_\text{coll} = 1 - \exp\biggl[ -\sigma_{gg} \int dx \, dy \int dz \, \rho_A \int dz \, \rho_B \biggr],
  \label{eq:pcoll}
\end{equation}
where the integral in the exponential is the overlap integral of the proton thickness functions and
$\sigma_{gg}$ is an effective parton-parton cross-section tuned so that the total proton-proton
cross-section equals the experimental inelastic nucleon-nucleon cross-section $\sigma_\text{NN}$.

\begin{figure*}[t]
  \includegraphics{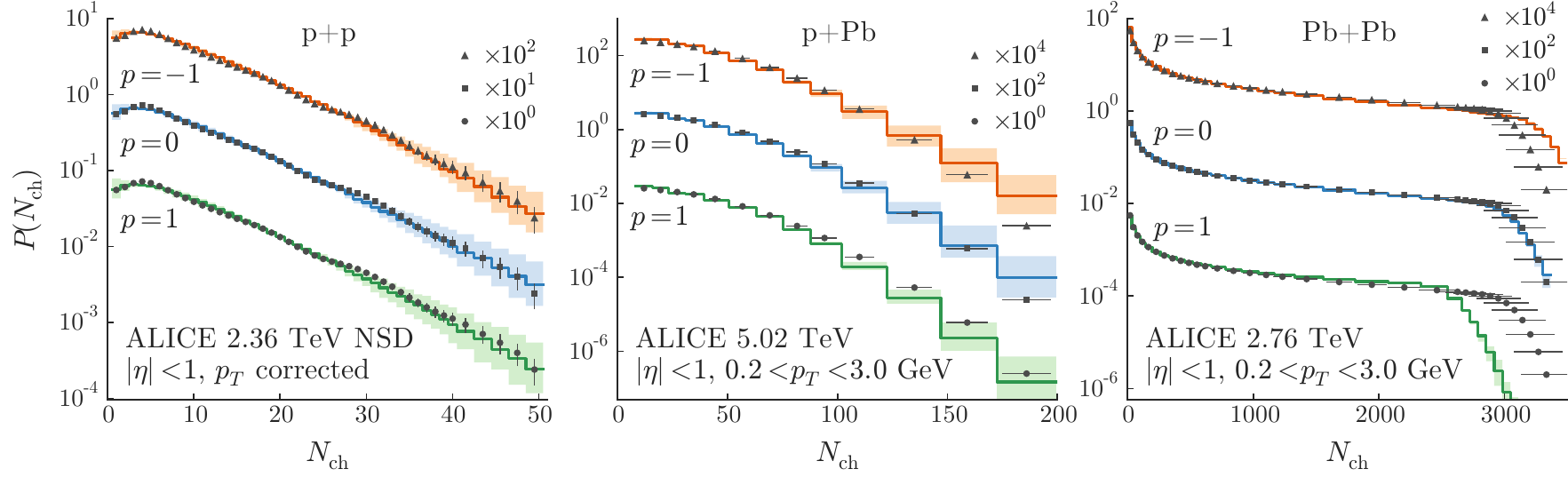}
  \caption{
    \label{fig:nch}
    Multiplicity distributions for proton-proton, proton-lead, and lead-lead collisions.
    The histograms are \protect\trento\ results for reduced thickness parameter $p = -1$ (top, orange), $p = 0$ (middle, blue), and $p = 1$ (bottom, green), with approximate best-fit fluctuation parameters $k$ and normalizations given in table~\ref{tab:nch}.
    The shaded bands show the sensitivity from varying $k$ by $\pm30\%$.
    Data points (triangles, squares, circles) are experimental distributions from ALICE \cite{Aamodt:2010ft,Abelev:2014mda} offset by powers of ten for comparison with the model.
  }
\end{figure*}

The collision probability is sampled once to determine if the protons collide; assuming they do, we follow a procedure similar to \cite{Bozek:2013uha} and assign each proton a \emph{fluctuated} thickness
\begin{equation}
  T_{A,B}(x, y) = w_{A,B} \int dz \, \rho_{A,B}(x, y, z),
\end{equation}
where $w_{A,B}$ are independent random weights sampled from a gamma distribution with unit mean,
\begin{equation}
  P_k(w) = \frac{k^k}{\Gamma(k)} w^{k-1} e^{-kw}.
  \label{eq:gamma}
\end{equation}
These gamma weights introduce additional multiplicity fluctuations in order to reproduce the large fluctuations observed in experimental proton-proton collisions.
The shape parameter $k$ may be tuned to optimally fit the data:
small values ($0 < k < 1$) correspond to large multiplicity fluctuations, while large values ($k \gg 1$) suppress fluctuations.

With the projectile thickness functions in hand, the reduced thickness is calculated to furnish the initial transverse entropy profile up to an overall normalization factor,
\begin{equation}
  dS/dy \, |_{\tau = \tau_0} \propto T_R(p; T_A, T_B).
  \label{eq:dSdy}
\end{equation}

\begin{table}[t]
  \caption{
    \label{tab:nch}
    Approximate best-fit fluctuation parameters $k$ and normalizations for each $p$ value and collision system in Fig.~\ref{fig:nch}.
  }
  \begin{ruledtabular}
  \begin{tabular}{rcccc}
    $p\;$  & $k$ & p+p norm & p+Pb norm & Pb+Pb norm \\
    \noalign{\smallskip}\hline\noalign{\smallskip}
    $+1$   & 0.8 & 9.7      & 7.0       & 13.        \\
    $ 0$   & 1.4 & 19.      & 17.       & 16.        \\
    $-1$   & 2.2 & 24.      & 26.       & 18.        \\
  \end{tabular}
  \end{ruledtabular}
\end{table}

Composite collision systems such as proton-nucleus and nucleus-nucleus are essentially treated as
superpositions of proton-proton collisions.
A set of nucleon positions is chosen for each projectile, typically by sampling an uncorrelated Woods-Saxon distribution or from more realistic correlated nuclear configurations when available \cite{Alvioli:2009ab}.
The collision probability \eqref{eq:pcoll} is sampled for each pairwise interaction and those nucleons that collide with at least one partner are labeled ``participants'' while the rest are discarded.
The fluctuated thickness function of nucleus $A$ then reads
\begin{equation}
  T_A = \sum_{i=1}^{N_\text{part}} w_i \int dz \, \rho_\text{proton}(x - x_i, y - y_i, z - z_i),
  \label{eq:nuc-thickness}
\end{equation}
where $w_i$ and $(x_i, y_i, z_i)$ are the weights and position, respectively, of participant $i$ in nucleus $A$.
$T_B$ follows analogously.

This completes the construction of the model, \trento\ (Reduced Thickness Event-by-event Nuclear Topology).
In summary, the model deposits entropy proportional to the reduced thickness function \eqref{eq:tr}, defined as the generalized mean of fluctuated participant thickness functions \eqref{eq:nuc-thickness}, with each participant nucleon weighted by an independent gamma random number \eqref{eq:gamma}.

\psec{Applications}

We now demonstrate \trento's ability to simultaneously describe a wide range of collision systems.
Note that the reduced thickness parameter $p$, gamma fluctuation parameter $k$, and nucleon profile $\rho_\text{proton}$ are not rigorously constrained---to do so would require a systematic model-to-data comparison \cite{Bernhard:2015hxa} which is beyond the scope of this work.
Therefore, the following results do not necessarily represent the best-fit of the model to data.

\begin{figure*}[t]
  \includegraphics{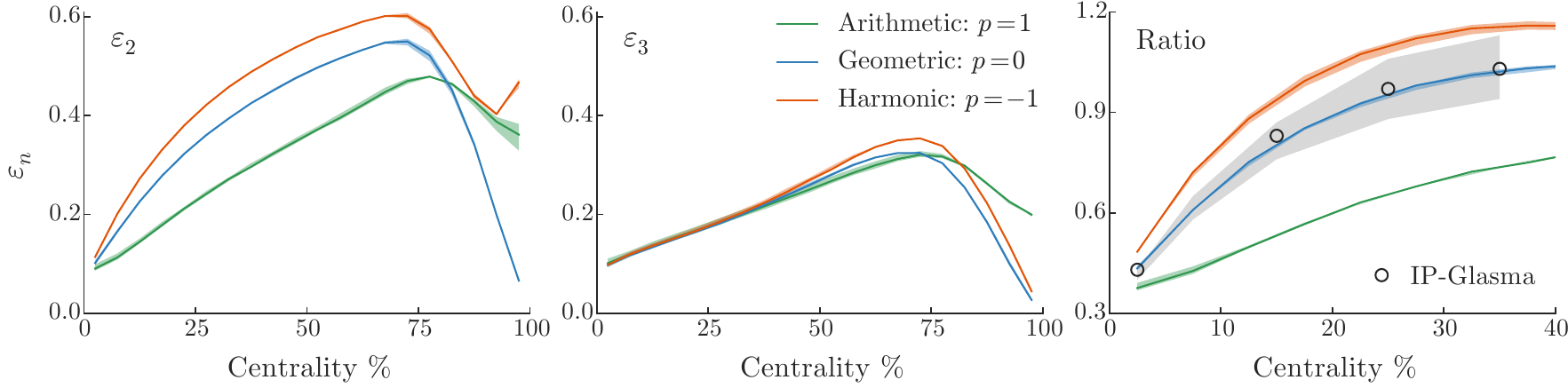}
  \caption{
    \label{fig:eccen}
    Left and middle plots:
    Eccentricity harmonics $\varepsilon_2$ and $\varepsilon_3$ as a function of centrality for reduced thickness parameters $p = 1$, 0, $-1$ (green, blue, orange).
    The shaded bands show the sensitivity from varying $k$ by $\pm30\%$ from the values in table~\ref{tab:nch}.
    Right plot:
    Ratio of the rms eccentricities $\eccratio$ against the allowed region (grey band) and the ratio computed by IP-Glasma (circles) \cite{Retinskaya:2013gca}.
    Note that the axes have different ranges in the ratio plot.
  }
\end{figure*}

We adopt a three-stage model for particle production similar to \cite{Bozek:2013uha}, in which the final multiplicity arises from a convolution of the initial entropy deposited by the collision, viscous entropy production during hydrodynamic evolution, and statistical hadronization at freeze-out.
The average charged-particle multiplicity $\avg\nch$ after hydrodynamic evolution is to a good approximation proportional to the total initial entropy \cite{Song:2008si} and hence to the integrated reduced thickness via Eq.~\eqref{eq:dSdy}:
\begin{equation}
  \avg\nch \propto \int dx \, dy \, T_R.
\end{equation}
Then, assuming independent particle emission at freeze-out, the final number of charged particles is Poisson distributed \cite{Kisiel:2005hn,*Chojnacki:2011hb}, i.e.~$P(\nch) = \text{Poisson}(\avg\nch)$.
The folding of the Poisson fluctuations with the gamma weights for each participant yields a negative binomial distribution \cite{Bozek:2013uha}, which has historically been used to fit proton-proton multiplicity fluctuations.

To compare with experimental multiplicity distributions, we generate a large ensemble of minimum-bias events, integrate their $T_R$ profiles, rescale by an overall normalization constant, and sample a Poisson number for the multiplicity of each event.
The left panel of Fig.~\ref{fig:nch} shows the $\nch$ distributions for proton-proton simulations with reduced thickness parameter $p = 1$, 0, $-1$, and Gaussian beam-integrated proton density
\begin{equation}
  \int dz \, \rho_\text{proton} = \frac{1}{2\pi B} \exp\biggr( -\frac{x^2 + y^2}{2B} \biggr)
\end{equation}
with effective area $B = (0.6\;\text{fm})^2$.
We tune the fluctuation parameter $k$ for each value of $p$ to qualitatively fit the experimental proton-proton distribution \cite{Aamodt:2010ft}, and additionally vary $k$ by $\pm30\%$ to explore the sensitivity of the model to the gamma participant weights.
For proton-lead and lead-lead collisions \cite{Abelev:2014mda} (middle and right panels), we use identical model parameters except for the overall normalization factor, which is allowed to vary independently across collision systems to account for differences in beam energy and kinematic cuts (annotated in the figure).
The $k$ values and normalizations are given in table~\ref{tab:nch}.

The model is able to reproduce the experimental proton-proton distribution for each value of $p$, provided $k$ is appropriately tuned.
Varying the best-fit $k$ value (by $\pm30\%$) has a noticeable effect on proton-proton and proton-lead systems, especially in the high-multiplicity tails, but is less important in lead-lead collisions, where the gamma weights are averaged over many participant nucleons.

Each $p$ value also yields a reasonable fit to the shapes of the proton-lead and lead-lead distributions, although lead-lead appears to favor $p \approx 0$.
Note that the normalizations for $p = 1$ (wounded nucleon model) in proton-lead and lead-lead collisions (table~\ref{tab:nch}) are not self-consistent, since proton-lead requires roughly half the normalization as lead-lead, even though the experimental data were measured at a higher beam energy.

Eccentricity harmonics $\varepsilon_n$ are calculated using the definition
\begin{equation}
  \varepsilon_n e^{i n\phi} = -\frac{\int dx \, dy\, r^n e^{i n \phi} \, T_R}{\int dx \, dy \, r^n \, T_R}.
\end{equation}
Figure~\ref{fig:eccen} shows ellipticity $\varepsilon_2$ and triangularity $\varepsilon_3$ as a function of centrality using the same lead-lead data as in Fig.~\ref{fig:nch}.
There is a clear trend of increasing eccentricity (particularly $\varepsilon_2$) with decreasing $p$.
This is a larger-scale manifestation of the behavior in Fig.~\ref{fig:TR}:
as $p$ decreases, the generalized mean \eqref{eq:tr} attenuates entropy production in asymmetric regions of the collision, accentuating the elliptical overlap shape in non-central collisions and enhancing their eccentricity.
Meanwhile, varying the fluctuation parameter $k$ has limited effect.

In addition, we perform the test proposed by \cite{Retinskaya:2013gca}, which uses flow data and hydrodynamic calculations to determine an experimentally allowed band for the ratio of root-mean-square eccentricities $\eccratio$ as a function of centrality.
Among available initial condition models only IP-Glasma consistently falls within the allowed region.
As shown in the right panel of Fig.~\ref{fig:eccen}, \trento\ with $p = 0$ (geometric mean) yields excellent agreement with the allowed band and is similar to IP-Glasma.

\begin{figure}[b]
  \tikzsetnextfilename{uranium_diagram}
  \begin{tikzpicture}[
    uranium/.style={draw, semithick, ellipse, anchor=center},
    small width/.style={minimum width=17},
    large width/.style={minimum width=31},
    small height/.style={minimum height=17},
    large height/.style={minimum height=31}
  ]
    \matrix (m) [matrix of nodes] {
      &[-2.1ex] Side view &[-2.1ex] & Beam view & $\varepsilon_2$ & $N_\text{part}$ & $N_\text{coll}$ \\[.7ex]
      |[uranium, large width, small height] (ttl)| U &
      |[above]| tip-tip &
      |[uranium, large width, small height] (ttr)| U &
      \node[draw, circle, small height, small width, xshift=1mm] {};
      \node[uranium, circle, small width, small height, fill=white] {U}; &
      smaller & equal & larger \\[1.7ex]
      |[uranium, large height, small width] (ssl)| U &
      |[above]| side-side &
      |[uranium, large height, small width] (ssr)| U &
      \node[draw, ellipse, large height, small width, xshift=1mm] {};
      \node[uranium, large height, small width, fill=white] {U}; &
      larger & equal & smaller \\
    };
    \begin{scope}[->]
      \draw (ttl) -- ($(ttl)!.48!(ttr)$);
      \draw (ttr) -- ($(ttr)!.48!(ttl)$);
      \draw (ssl) -- ($(ssl)!.48!(ssr)$);
      \draw (ssr) -- ($(ssr)!.48!(ssl)$);
    \end{scope}
  \end{tikzpicture}
  \caption{
    \label{fig:uu-schematic}
    Comparison of tip-tip and side-side uranium-uranium collisions.
    Schematics are shown from a side view and looking down the beam axis, and the following quantities are compared:
    ellipticity $\varepsilon_2$, number of participating nucleons $N_\text{part}$, and number of binary nucleon-nucleon collisions $N_\text{coll}$.
  }
\end{figure}
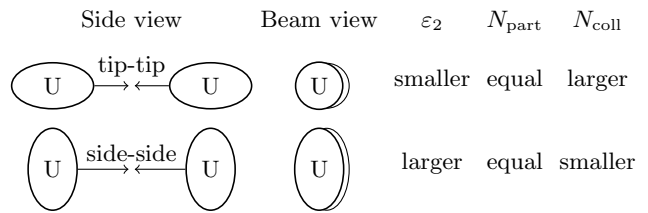

As a final novel application, we return to the previously mentioned ultra-central uranium-uranium puzzle, where typical Glauber models are notably inconsistent with experimental data.
Unlike e.g.~gold and lead, uranium nuclei have a highly deformed prolate spheroidal shape, so uranium-uranium collisions may achieve maximal overlap via two distinct orientations:
``tip-tip'', in which the long axes of the spheroids are aligned with the beam axis and the overlap area is circular;
or ``side-side'', where the long axes are perpendicular to the beam axis and the overlap area is elliptical, as shown in Fig.~\ref{fig:uu-schematic}.
Hence side-side collisions will in general have larger initial-state ellipticity $\varepsilon_2$ and final-state elliptic flow $v_2$ than tip-tip.

In the two-component Glauber model, tip-tip collisions produce more binary nucleon-nucleon collisions than side-side, so tip-tip collisions have larger charged-particle multiplicity $\nch$.
Therefore, the most central uranium-uranium events are dominated by tip-tip collisions with maximal $\nch$ and small $v_2$, while side-side collisions have a smaller $\nch$ and somewhat larger $v_2$.
This predicted drop in elliptic flow as a function of $\nch$ is known as the ``knee'' \cite{Voloshin:2010ut}.

\begin{figure}[t]
  \centering
  \includegraphics{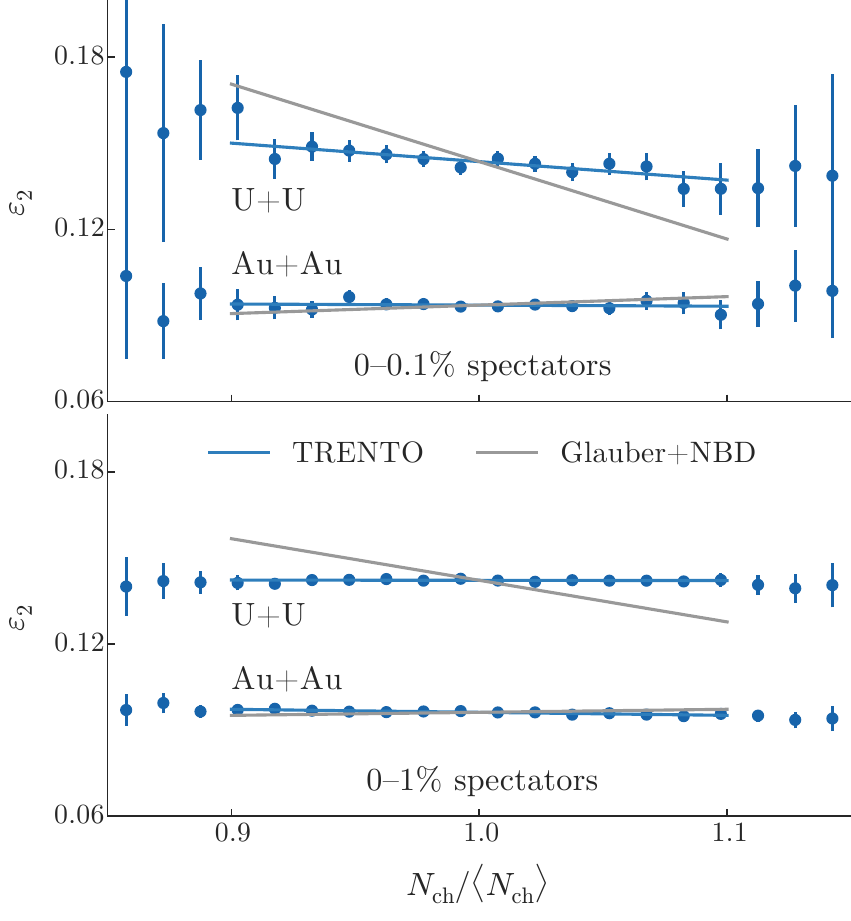}
  \caption{
    \label{fig:uranium}
    Ellipticity $\varepsilon_2$ as a function of normalized charged-particle multiplicity $\nch/\avg\nch$ in ultra-central uranium-uranium and gold-gold collisions at RHIC.
    The top and bottom plots show the top 0.1\% and 1\% of collisions selected by number of spectators to mimic STAR's experimental ZDC selection \cite{FortheSTAR:2013bza}.
    Blue points with error bars are binned \protect\trento\ results with reduced thickness parameter $p = 0$ and best-fit fluctuation parameter $k = 1.4$.
    Blue lines are linear fits within $0.9~<~\nch/\avg\nch~<~1.1$.
    Grey lines represent the analogous Glauber+NBD slopes calculated in \cite{FortheSTAR:2013bza}.
  }
\end{figure}

Recent data by STAR on uranium-uranium collisions exhibits no evidence of a knee \cite{FortheSTAR:2013bza,Wang:2014qxa}, at odds with Glauber model predictions.
It has been proposed that fluctuations could wash out the knee \cite{Rybczynski:2012av}, but a recent flow analysis showed that it would still be visible \cite{Goldschmidt:2015qya}.

The data therefore imply that multiplicity is independent of the number of binary collisions, justifying the scale-invariant condition \eqref{eq:scale-inv} postulated during the construction of the reduced thickness ansatz \eqref{eq:tr}.
Consequently, \trento\ predicts roughly the same number of charged particles in tip-tip and side-side uranium-uranium collisions.
As shown in Fig.~\ref{fig:uranium}, the slope of $\varepsilon_2$ as a function of $\nch$ is approximately equal for uranium-uranium and gold-gold, in contrast to the Glauber model which predicts a much steeper slope for uranium.
Short of conducting a full hydrodynamic analysis, \trento\ appears to be more consistent with STAR data than the Glauber model, and behaves similarly to IP-Glasma \cite{Schenke:2014tga}.

\psec{Conclusion}

In summary, we have developed \trento, a new parametric initial condition model for high-energy nuclear collisions that deposits entropy proportional to the generalized mean of nuclear overlap density.
This functional form smoothly interpolates among a family of physically reasonable models, enabling an unbiased model-to-data comparison to constrain the characteristics of the initially formed highly excited QCD matter.

We explored several discrete combinations of model parameters and found that entropy deposition functionally equivalent to the geometric mean simultaneously describes proton-proton, proton-lead, and lead-lead multiplicity distributions at the LHC and further satisfies experimentally extracted flow constraints on lead-lead eccentricity harmonics.
In addition, we showed that the model is qualitatively compatible with ultra-central uranium-uranium data, unlike the two-component Monte Carlo Glauber model with a binary collision term.

As a next step, we will couple the initial condition generator to viscous relativistic fluid dynamics as part of a complete simulation of the spacetime evolution of relativistic heavy-ion collisions \cite{Shen:2014vra} and embed the evolution model in a systematic model-to-data comparison framework \cite{Bernhard:2015hxa}.
This will enable simultaneous extraction of initial condition parameters and sQGP medium properties with quantitative uncertainties.

\medskip
\trento\ is publicly available at \url{github.com/Duke-QCD/trento}.
The community is encouraged to use, test, and contribute to the source code.

\psec{Acknowledgements}

\medskip
We would like to thank Ulrich Heinz, Scott Pratt, Ron Soltz, and Berndt M\"uller for helpful discussions and valuable feedback.
SAB is being supported by the U.S.\ Department of Energy Grant no.~DE-FG02-05ER41367,
JEB is supported through NSF grant no.~PHY-0941373,
and JSM acknowledges support by the DOE/NNSA Stockpile Stewardship Graduate Fellowship under grant no.~DE-FC52-08NA28752.

\bibliography{trento,duke-qcd-refs/Duke_QCD_refs}

\begin{thebibliography}{44}%
\makeatletter
\providecommand \@ifxundefined [1]{%
 \@ifx{#1\undefined}
}%
\providecommand \@ifnum [1]{%
 \ifnum #1\expandafter \@firstoftwo
 \else \expandafter \@secondoftwo
 \fi
}%
\providecommand \@ifx [1]{%
 \ifx #1\expandafter \@firstoftwo
 \else \expandafter \@secondoftwo
 \fi
}%
\providecommand \natexlab [1]{#1}%
\providecommand \enquote  [1]{``#1''}%
\providecommand \bibnamefont  [1]{#1}%
\providecommand \bibfnamefont [1]{#1}%
\providecommand \citenamefont [1]{#1}%
\providecommand \href@noop [0]{\@secondoftwo}%
\providecommand \href [0]{\begingroup \@sanitize@url \@href}%
\providecommand \@href[1]{\@@startlink{#1}\@@href}%
\providecommand \@@href[1]{\endgroup#1\@@endlink}%
\providecommand \@sanitize@url [0]{\catcode `\\12\catcode `\$12\catcode
  `\&12\catcode `\#12\catcode `\^12\catcode `\_12\catcode `\%12\relax}%
\providecommand \@@startlink[1]{}%
\providecommand \@@endlink[0]{}%
\providecommand \url  [0]{\begingroup\@sanitize@url \@url }%
\providecommand \@url [1]{\endgroup\@href {#1}{\urlprefix }}%
\providecommand \urlprefix  [0]{URL }%
\providecommand \Eprint [0]{\href }%
\providecommand \doibase [0]{http://dx.doi.org/}%
\providecommand \selectlanguage [0]{\@gobble}%
\providecommand \bibinfo  [0]{\@secondoftwo}%
\providecommand \bibfield  [0]{\@secondoftwo}%
\providecommand \translation [1]{[#1]}%
\providecommand \BibitemOpen [0]{}%
\providecommand \bibitemStop [0]{}%
\providecommand \bibitemNoStop [0]{.\EOS\space}%
\providecommand \EOS [0]{\spacefactor3000\relax}%
\providecommand \BibitemShut  [1]{\csname bibitem#1\endcsname}%
\let\auto@bib@innerbib\@empty
\bibitem [{\citenamefont {Arsene}\ \emph {et~al.}(2005)\citenamefont {Arsene}
  \emph {et~al.}}]{Arsene:2004fa}%
  \BibitemOpen
  \bibfield  {author} {\bibinfo {author} {\bibfnamefont {I.}~\bibnamefont
  {Arsene}} \emph {et~al.} (\bibinfo {collaboration} {BRAHMS}),\ }\href@noop {}
  {\bibfield  {journal} {\bibinfo  {journal} {Nucl. Phys.}\ }\textbf {\bibinfo
  {volume} {A757}},\ \bibinfo {pages} {1} (\bibinfo {year} {2005})},\ \Eprint
  {http://arxiv.org/abs/nucl-ex/0410020} {nucl-ex/0410020} \BibitemShut
  {NoStop}%
\bibitem [{\citenamefont {Adcox}\ \emph {et~al.}(2005)\citenamefont {Adcox}
  \emph {et~al.}}]{Adcox:2004mh}%
  \BibitemOpen
  \bibfield  {author} {\bibinfo {author} {\bibfnamefont {K.}~\bibnamefont
  {Adcox}} \emph {et~al.} (\bibinfo {collaboration} {PHENIX}),\ }\href@noop {}
  {\bibfield  {journal} {\bibinfo  {journal} {Nucl. Phys.}\ }\textbf {\bibinfo
  {volume} {A757}},\ \bibinfo {pages} {184} (\bibinfo {year} {2005})},\ \Eprint
  {http://arxiv.org/abs/nucl-ex/0410003} {nucl-ex/0410003} \BibitemShut
  {NoStop}%
\bibitem [{\citenamefont {Back}\ \emph {et~al.}(2005)\citenamefont {Back} \emph
  {et~al.}}]{Back:2004je}%
  \BibitemOpen
  \bibfield  {author} {\bibinfo {author} {\bibfnamefont {B.~B.}\ \bibnamefont
  {Back}} \emph {et~al.},\ }\href@noop {} {\bibfield  {journal} {\bibinfo
  {journal} {Nucl. Phys.}\ }\textbf {\bibinfo {volume} {A757}},\ \bibinfo
  {pages} {28} (\bibinfo {year} {2005})},\ \Eprint
  {http://arxiv.org/abs/nucl-ex/0410022} {nucl-ex/0410022} \BibitemShut
  {NoStop}%
\bibitem [{\citenamefont {Adams}\ \emph {et~al.}(2005)\citenamefont {Adams}
  \emph {et~al.}}]{Adams:2005dq}%
  \BibitemOpen
  \bibfield  {author} {\bibinfo {author} {\bibfnamefont {J.}~\bibnamefont
  {Adams}} \emph {et~al.} (\bibinfo {collaboration} {STAR}),\ }\href@noop {}
  {\bibfield  {journal} {\bibinfo  {journal} {Nucl. Phys.}\ }\textbf {\bibinfo
  {volume} {A757}},\ \bibinfo {pages} {102} (\bibinfo {year} {2005})},\ \Eprint
  {http://arxiv.org/abs/nucl-ex/0501009} {nucl-ex/0501009} \BibitemShut
  {NoStop}%
\bibitem [{\citenamefont {Gyulassy}\ and\ \citenamefont
  {McLerran}(2005)}]{Gyulassy:2004zy}%
  \BibitemOpen
  \bibfield  {author} {\bibinfo {author} {\bibfnamefont {M.}~\bibnamefont
  {Gyulassy}}\ and\ \bibinfo {author} {\bibfnamefont {L.}~\bibnamefont
  {McLerran}},\ }\href@noop {} {\bibfield  {journal} {\bibinfo  {journal}
  {Nucl. Phys.}\ }\textbf {\bibinfo {volume} {A750}},\ \bibinfo {pages} {30}
  (\bibinfo {year} {2005})},\ \Eprint {http://arxiv.org/abs/nucl-th/0405013}
  {nucl-th/0405013} \BibitemShut {NoStop}%
\bibitem [{\citenamefont {Muller}\ and\ \citenamefont
  {Nagle}(2006)}]{Muller:2006ee}%
  \BibitemOpen
  \bibfield  {author} {\bibinfo {author} {\bibfnamefont {B.}~\bibnamefont
  {Muller}}\ and\ \bibinfo {author} {\bibfnamefont {J.~L.}\ \bibnamefont
  {Nagle}},\ }\href {\doibase 10.1146/annurev.nucl.56.080805.140556} {\bibfield
   {journal} {\bibinfo  {journal} {Ann. Rev. Nucl. Part. Sci.}\ }\textbf
  {\bibinfo {volume} {56}},\ \bibinfo {pages} {93} (\bibinfo {year} {2006})},\
  \Eprint {http://arxiv.org/abs/nucl-th/0602029} {arXiv:nucl-th/0602029}
  \BibitemShut {NoStop}%
\bibitem [{\citenamefont {Muller}\ \emph {et~al.}(2012)\citenamefont {Muller},
  \citenamefont {Schukraft},\ and\ \citenamefont {Wyslouch}}]{Muller:2012zq}%
  \BibitemOpen
  \bibfield  {author} {\bibinfo {author} {\bibfnamefont {B.}~\bibnamefont
  {Muller}}, \bibinfo {author} {\bibfnamefont {J.}~\bibnamefont {Schukraft}}, \
  and\ \bibinfo {author} {\bibfnamefont {B.}~\bibnamefont {Wyslouch}},\ }\href
  {\doibase 10.1146/annurev-nucl-102711-094910} {\bibfield  {journal} {\bibinfo
   {journal} {Ann.Rev.Nucl.Part.Sci.}\ }\textbf {\bibinfo {volume} {62}},\
  \bibinfo {pages} {361} (\bibinfo {year} {2012})},\ \Eprint
  {http://arxiv.org/abs/1202.3233} {arXiv:1202.3233 [hep-ex]} \BibitemShut
  {NoStop}%
\bibitem [{\citenamefont {Petersen}\ \emph {et~al.}(2011)\citenamefont
  {Petersen}, \citenamefont {Coleman-Smith}, \citenamefont {Bass},\ and\
  \citenamefont {Wolpert}}]{Petersen:2010zt}%
  \BibitemOpen
  \bibfield  {author} {\bibinfo {author} {\bibfnamefont {H.}~\bibnamefont
  {Petersen}}, \bibinfo {author} {\bibfnamefont {C.}~\bibnamefont
  {Coleman-Smith}}, \bibinfo {author} {\bibfnamefont {S.~A.}\ \bibnamefont
  {Bass}}, \ and\ \bibinfo {author} {\bibfnamefont {R.}~\bibnamefont
  {Wolpert}},\ }\href {\doibase 10.1088/0954-3899/38/4/045102} {\bibfield
  {journal} {\bibinfo  {journal} {J.Phys.G}\ }\textbf {\bibinfo {volume}
  {G38}},\ \bibinfo {pages} {045102} (\bibinfo {year} {2011})},\ \Eprint
  {http://arxiv.org/abs/1012.4629} {arXiv:1012.4629 [nucl-th]} \BibitemShut
  {NoStop}%
\bibitem [{\citenamefont {Novak}\ \emph {et~al.}(2014)\citenamefont {Novak},
  \citenamefont {Novak}, \citenamefont {Pratt}, \citenamefont {Vredevoogd},
  \citenamefont {Coleman-Smith},\ and\ \citenamefont
  {Wolpert}}]{Novak:2013bqa}%
  \BibitemOpen
  \bibfield  {author} {\bibinfo {author} {\bibfnamefont {J.}~\bibnamefont
  {Novak}}, \bibinfo {author} {\bibfnamefont {K.}~\bibnamefont {Novak}},
  \bibinfo {author} {\bibfnamefont {S.}~\bibnamefont {Pratt}}, \bibinfo
  {author} {\bibfnamefont {J.}~\bibnamefont {Vredevoogd}}, \bibinfo {author}
  {\bibfnamefont {C.}~\bibnamefont {Coleman-Smith}}, \ and\ \bibinfo {author}
  {\bibfnamefont {R.}~\bibnamefont {Wolpert}},\ }\href {\doibase
  10.1103/PhysRevC.89.034917} {\bibfield  {journal} {\bibinfo  {journal}
  {Phys.Rev.}\ }\textbf {\bibinfo {volume} {C89}},\ \bibinfo {pages} {034917}
  (\bibinfo {year} {2014})},\ \Eprint {http://arxiv.org/abs/1303.5769}
  {arXiv:1303.5769 [nucl-th]} \BibitemShut {NoStop}%
\bibitem [{\citenamefont {Baier}\ and\ \citenamefont
  {Romatschke}(2007)}]{Baier:2006gy}%
  \BibitemOpen
  \bibfield  {author} {\bibinfo {author} {\bibfnamefont {R.}~\bibnamefont
  {Baier}}\ and\ \bibinfo {author} {\bibfnamefont {P.}~\bibnamefont
  {Romatschke}},\ }\href {\doibase 10.1140/epjc/s10052-007-0308-5} {\bibfield
  {journal} {\bibinfo  {journal} {Eur.Phys.J.}\ }\textbf {\bibinfo {volume}
  {C51}},\ \bibinfo {pages} {677} (\bibinfo {year} {2007})},\ \Eprint
  {http://arxiv.org/abs/nucl-th/0610108} {arXiv:nucl-th/0610108 [nucl-th]}
  \BibitemShut {NoStop}%
\bibitem [{\citenamefont {Song}\ and\ \citenamefont
  {Heinz}(2008{\natexlab{a}})}]{Song:2007ux}%
  \BibitemOpen
  \bibfield  {author} {\bibinfo {author} {\bibfnamefont {H.}~\bibnamefont
  {Song}}\ and\ \bibinfo {author} {\bibfnamefont {U.~W.}\ \bibnamefont
  {Heinz}},\ }\href {\doibase 10.1103/PhysRevC.77.064901} {\bibfield  {journal}
  {\bibinfo  {journal} {Phys. Rev.}\ }\textbf {\bibinfo {volume} {C77}},\
  \bibinfo {pages} {064901} (\bibinfo {year} {2008}{\natexlab{a}})},\ \Eprint
  {http://arxiv.org/abs/0712.3715} {arXiv:0712.3715 [nucl-th]} \BibitemShut
  {NoStop}%
\bibitem [{\citenamefont {Luzum}\ and\ \citenamefont
  {Romatschke}(2008)}]{Luzum:2008cw}%
  \BibitemOpen
  \bibfield  {author} {\bibinfo {author} {\bibfnamefont {M.}~\bibnamefont
  {Luzum}}\ and\ \bibinfo {author} {\bibfnamefont {P.}~\bibnamefont
  {Romatschke}},\ }\href {\doibase 10.1103/PhysRevC.78.034915} {\bibfield
  {journal} {\bibinfo  {journal} {Phys. Rev.}\ }\textbf {\bibinfo {volume}
  {C78}},\ \bibinfo {pages} {034915} (\bibinfo {year} {2008})},\ \Eprint
  {http://arxiv.org/abs/0804.4015} {arXiv:0804.4015 [nucl-th]} \BibitemShut
  {NoStop}%
\bibitem [{\citenamefont {Schenke}\ \emph {et~al.}(2011)\citenamefont
  {Schenke}, \citenamefont {Jeon},\ and\ \citenamefont
  {Gale}}]{Schenke:2010rr}%
  \BibitemOpen
  \bibfield  {author} {\bibinfo {author} {\bibfnamefont {B.}~\bibnamefont
  {Schenke}}, \bibinfo {author} {\bibfnamefont {S.}~\bibnamefont {Jeon}}, \
  and\ \bibinfo {author} {\bibfnamefont {C.}~\bibnamefont {Gale}},\ }\href
  {\doibase 10.1103/PhysRevLett.106.042301} {\bibfield  {journal} {\bibinfo
  {journal} {Phys.Rev.Lett.}\ }\textbf {\bibinfo {volume} {106}},\ \bibinfo
  {pages} {042301} (\bibinfo {year} {2011})},\ \Eprint
  {http://arxiv.org/abs/1009.3244} {arXiv:1009.3244 [hep-ph]} \BibitemShut
  {NoStop}%
\bibitem [{\citenamefont {Shen}\ \emph {et~al.}(2011)\citenamefont {Shen},
  \citenamefont {Heinz}, \citenamefont {Huovinen},\ and\ \citenamefont
  {Song}}]{Shen:2011eg}%
  \BibitemOpen
  \bibfield  {author} {\bibinfo {author} {\bibfnamefont {C.}~\bibnamefont
  {Shen}}, \bibinfo {author} {\bibfnamefont {U.}~\bibnamefont {Heinz}},
  \bibinfo {author} {\bibfnamefont {P.}~\bibnamefont {Huovinen}}, \ and\
  \bibinfo {author} {\bibfnamefont {H.}~\bibnamefont {Song}},\ }\href {\doibase
  10.1103/PhysRevC.84.044903} {\bibfield  {journal} {\bibinfo  {journal}
  {Phys.Rev.}\ }\textbf {\bibinfo {volume} {C84}},\ \bibinfo {pages} {044903}
  (\bibinfo {year} {2011})},\ \Eprint {http://arxiv.org/abs/1105.3226}
  {arXiv:1105.3226 [nucl-th]} \BibitemShut {NoStop}%
\bibitem [{\citenamefont {Shen}\ \emph {et~al.}(2014)\citenamefont {Shen},
  \citenamefont {Qiu}, \citenamefont {Song}, \citenamefont {Bernhard},
  \citenamefont {Bass},\ and\ \citenamefont {Heinz}}]{Shen:2014vra}%
  \BibitemOpen
  \bibfield  {author} {\bibinfo {author} {\bibfnamefont {C.}~\bibnamefont
  {Shen}}, \bibinfo {author} {\bibfnamefont {Z.}~\bibnamefont {Qiu}}, \bibinfo
  {author} {\bibfnamefont {H.}~\bibnamefont {Song}}, \bibinfo {author}
  {\bibfnamefont {J.}~\bibnamefont {Bernhard}}, \bibinfo {author}
  {\bibfnamefont {S.}~\bibnamefont {Bass}}, \ and\ \bibinfo {author}
  {\bibfnamefont {U.}~\bibnamefont {Heinz}},\ }\href@noop {} {\  (\bibinfo
  {year} {2014})},\ \Eprint {http://arxiv.org/abs/1409.8164} {arXiv:1409.8164
  [nucl-th]} \BibitemShut {NoStop}%
\bibitem [{\citenamefont {Song}\ \emph {et~al.}(2011)\citenamefont {Song},
  \citenamefont {Bass}, \citenamefont {Heinz}, \citenamefont {Hirano},\ and\
  \citenamefont {Shen}}]{Song:2010mg}%
  \BibitemOpen
  \bibfield  {author} {\bibinfo {author} {\bibfnamefont {H.}~\bibnamefont
  {Song}}, \bibinfo {author} {\bibfnamefont {S.~A.}\ \bibnamefont {Bass}},
  \bibinfo {author} {\bibfnamefont {U.}~\bibnamefont {Heinz}}, \bibinfo
  {author} {\bibfnamefont {T.}~\bibnamefont {Hirano}}, \ and\ \bibinfo {author}
  {\bibfnamefont {C.}~\bibnamefont {Shen}},\ }\href {\doibase
  10.1103/PhysRevLett.106.192301} {\bibfield  {journal} {\bibinfo  {journal}
  {Phys.Rev.Lett.}\ }\textbf {\bibinfo {volume} {106}},\ \bibinfo {pages}
  {192301} (\bibinfo {year} {2011})},\ \Eprint {http://arxiv.org/abs/1011.2783}
  {arXiv:1011.2783 [nucl-th]} \BibitemShut {NoStop}%
\bibitem [{\citenamefont {Retinskaya}\ \emph {et~al.}(2014)\citenamefont
  {Retinskaya}, \citenamefont {Luzum},\ and\ \citenamefont
  {Ollitrault}}]{Retinskaya:2013gca}%
  \BibitemOpen
  \bibfield  {author} {\bibinfo {author} {\bibfnamefont {E.}~\bibnamefont
  {Retinskaya}}, \bibinfo {author} {\bibfnamefont {M.}~\bibnamefont {Luzum}}, \
  and\ \bibinfo {author} {\bibfnamefont {J.-Y.}\ \bibnamefont {Ollitrault}},\
  }\href {\doibase 10.1103/PhysRevC.89.014902} {\bibfield  {journal} {\bibinfo
  {journal} {Phys.Rev.}\ }\textbf {\bibinfo {volume} {C89}},\ \bibinfo {pages}
  {014902} (\bibinfo {year} {2014})},\ \Eprint {http://arxiv.org/abs/1311.5339}
  {arXiv:1311.5339 [nucl-th]} \BibitemShut {NoStop}%
\bibitem [{\citenamefont {Schenke}\ \emph {et~al.}(2012)\citenamefont
  {Schenke}, \citenamefont {Tribedy},\ and\ \citenamefont
  {Venugopalan}}]{Schenke:2012wb}%
  \BibitemOpen
  \bibfield  {author} {\bibinfo {author} {\bibfnamefont {B.}~\bibnamefont
  {Schenke}}, \bibinfo {author} {\bibfnamefont {P.}~\bibnamefont {Tribedy}}, \
  and\ \bibinfo {author} {\bibfnamefont {R.}~\bibnamefont {Venugopalan}},\
  }\href {\doibase 10.1103/PhysRevLett.108.252301} {\bibfield  {journal}
  {\bibinfo  {journal} {Phys.Rev.Lett.}\ }\textbf {\bibinfo {volume} {108}},\
  \bibinfo {pages} {252301} (\bibinfo {year} {2012})},\ \Eprint
  {http://arxiv.org/abs/1202.6646} {arXiv:1202.6646 [nucl-th]} \BibitemShut
  {NoStop}%
\bibitem [{\citenamefont {van~der Schee}\ \emph {et~al.}(2013)\citenamefont
  {van~der Schee}, \citenamefont {Romatschke},\ and\ \citenamefont
  {Pratt}}]{vanderSchee:2013pia}%
  \BibitemOpen
  \bibfield  {author} {\bibinfo {author} {\bibfnamefont {W.}~\bibnamefont
  {van~der Schee}}, \bibinfo {author} {\bibfnamefont {P.}~\bibnamefont
  {Romatschke}}, \ and\ \bibinfo {author} {\bibfnamefont {S.}~\bibnamefont
  {Pratt}},\ }\href {\doibase 10.1103/PhysRevLett.111.222302} {\bibfield
  {journal} {\bibinfo  {journal} {Phys.Rev.Lett.}\ }\textbf {\bibinfo {volume}
  {111}},\ \bibinfo {pages} {222302} (\bibinfo {year} {2013})},\ \Eprint
  {http://arxiv.org/abs/1307.2539} {arXiv:1307.2539} \BibitemShut {NoStop}%
\bibitem [{\citenamefont {Berges}\ \emph {et~al.}(2014)\citenamefont {Berges},
  \citenamefont {Schenke}, \citenamefont {Schlichting},\ and\ \citenamefont
  {Venugopalan}}]{Berges:2014yta}%
  \BibitemOpen
  \bibfield  {author} {\bibinfo {author} {\bibfnamefont {J.}~\bibnamefont
  {Berges}}, \bibinfo {author} {\bibfnamefont {B.}~\bibnamefont {Schenke}},
  \bibinfo {author} {\bibfnamefont {S.}~\bibnamefont {Schlichting}}, \ and\
  \bibinfo {author} {\bibfnamefont {R.}~\bibnamefont {Venugopalan}},\ }\href
  {\doibase 10.1016/j.nuclphysa.2014.08.103} {\  (\bibinfo {year} {2014}),\
  10.1016/j.nuclphysa.2014.08.103},\ \Eprint {http://arxiv.org/abs/1409.1638}
  {arXiv:1409.1638 [hep-ph]} \BibitemShut {NoStop}%
\bibitem [{\citenamefont {Kurkela}\ and\ \citenamefont
  {Lu}(2014)}]{Kurkela:2014tea}%
  \BibitemOpen
  \bibfield  {author} {\bibinfo {author} {\bibfnamefont {A.}~\bibnamefont
  {Kurkela}}\ and\ \bibinfo {author} {\bibfnamefont {E.}~\bibnamefont {Lu}},\
  }\href {\doibase 10.1103/PhysRevLett.113.182301} {\bibfield  {journal}
  {\bibinfo  {journal} {Phys.Rev.Lett.}\ }\textbf {\bibinfo {volume} {113}},\
  \bibinfo {pages} {182301} (\bibinfo {year} {2014})},\ \Eprint
  {http://arxiv.org/abs/1405.6318} {arXiv:1405.6318 [hep-ph]} \BibitemShut
  {NoStop}%
\bibitem [{\citenamefont {McLerran}\ and\ \citenamefont
  {Venugopalan}(1994{\natexlab{a}})}]{McLerran:1993ni}%
  \BibitemOpen
  \bibfield  {author} {\bibinfo {author} {\bibfnamefont {L.~D.}\ \bibnamefont
  {McLerran}}\ and\ \bibinfo {author} {\bibfnamefont {R.}~\bibnamefont
  {Venugopalan}},\ }\href@noop {} {\bibfield  {journal} {\bibinfo  {journal}
  {Phys. Rev.}\ }\textbf {\bibinfo {volume} {D49}},\ \bibinfo {pages} {2233}
  (\bibinfo {year} {1994}{\natexlab{a}})},\ \Eprint
  {http://arxiv.org/abs/hep-ph/9309289} {hep-ph/9309289} \BibitemShut {NoStop}%
\bibitem [{\citenamefont {McLerran}\ and\ \citenamefont
  {Venugopalan}(1994{\natexlab{b}})}]{McLerran:1993ka}%
  \BibitemOpen
  \bibfield  {author} {\bibinfo {author} {\bibfnamefont {L.~D.}\ \bibnamefont
  {McLerran}}\ and\ \bibinfo {author} {\bibfnamefont {R.}~\bibnamefont
  {Venugopalan}},\ }\href@noop {} {\bibfield  {journal} {\bibinfo  {journal}
  {Phys. Rev.}\ }\textbf {\bibinfo {volume} {D49}},\ \bibinfo {pages} {3352}
  (\bibinfo {year} {1994}{\natexlab{b}})},\ \Eprint
  {http://arxiv.org/abs/hep-ph/9311205} {hep-ph/9311205} \BibitemShut {NoStop}%
\bibitem [{\citenamefont {Gelis}\ \emph {et~al.}(2010)\citenamefont {Gelis},
  \citenamefont {Iancu}, \citenamefont {Jalilian-Marian},\ and\ \citenamefont
  {Venugopalan}}]{Gelis:2010nm}%
  \BibitemOpen
  \bibfield  {author} {\bibinfo {author} {\bibfnamefont {F.}~\bibnamefont
  {Gelis}}, \bibinfo {author} {\bibfnamefont {E.}~\bibnamefont {Iancu}},
  \bibinfo {author} {\bibfnamefont {J.}~\bibnamefont {Jalilian-Marian}}, \ and\
  \bibinfo {author} {\bibfnamefont {R.}~\bibnamefont {Venugopalan}},\ }\href
  {\doibase 10.1146/annurev.nucl.010909.083629} {\bibfield  {journal} {\bibinfo
   {journal} {Ann.Rev.Nucl.Part.Sci.}\ }\textbf {\bibinfo {volume} {60}},\
  \bibinfo {pages} {463} (\bibinfo {year} {2010})},\ \Eprint
  {http://arxiv.org/abs/1002.0333} {arXiv:1002.0333 [hep-ph]} \BibitemShut
  {NoStop}%
\bibitem [{\citenamefont {Schenke}\ and\ \citenamefont
  {Venugopalan}(2014)}]{Schenke:2014zha}%
  \BibitemOpen
  \bibfield  {author} {\bibinfo {author} {\bibfnamefont {B.}~\bibnamefont
  {Schenke}}\ and\ \bibinfo {author} {\bibfnamefont {R.}~\bibnamefont
  {Venugopalan}},\ }\href {\doibase 10.1103/PhysRevLett.113.102301} {\bibfield
  {journal} {\bibinfo  {journal} {Phys.Rev.Lett.}\ }\textbf {\bibinfo {volume}
  {113}},\ \bibinfo {pages} {102301} (\bibinfo {year} {2014})},\ \Eprint
  {http://arxiv.org/abs/1405.3605} {arXiv:1405.3605 [nucl-th]} \BibitemShut
  {NoStop}%
\bibitem [{\citenamefont {Schenke}\ \emph
  {et~al.}(2014{\natexlab{a}})\citenamefont {Schenke}, \citenamefont
  {Tribedy},\ and\ \citenamefont {Venugopalan}}]{Schenke:2013dpa}%
  \BibitemOpen
  \bibfield  {author} {\bibinfo {author} {\bibfnamefont {B.}~\bibnamefont
  {Schenke}}, \bibinfo {author} {\bibfnamefont {P.}~\bibnamefont {Tribedy}}, \
  and\ \bibinfo {author} {\bibfnamefont {R.}~\bibnamefont {Venugopalan}},\
  }\href {\doibase 10.1103/PhysRevC.89.024901} {\bibfield  {journal} {\bibinfo
  {journal} {Phys.Rev.}\ }\textbf {\bibinfo {volume} {C89}},\ \bibinfo {pages}
  {024901} (\bibinfo {year} {2014}{\natexlab{a}})},\ \Eprint
  {http://arxiv.org/abs/1311.3636} {arXiv:1311.3636 [hep-ph]} \BibitemShut
  {NoStop}%
\bibitem [{\citenamefont {Miller}\ \emph {et~al.}(2007)\citenamefont {Miller},
  \citenamefont {Reygers}, \citenamefont {Sanders},\ and\ \citenamefont
  {Steinberg}}]{Miller:2007ri}%
  \BibitemOpen
  \bibfield  {author} {\bibinfo {author} {\bibfnamefont {M.~L.}\ \bibnamefont
  {Miller}}, \bibinfo {author} {\bibfnamefont {K.}~\bibnamefont {Reygers}},
  \bibinfo {author} {\bibfnamefont {S.~J.}\ \bibnamefont {Sanders}}, \ and\
  \bibinfo {author} {\bibfnamefont {P.}~\bibnamefont {Steinberg}},\ }\href
  {\doibase 10.1146/annurev.nucl.57.090506.123020} {\bibfield  {journal}
  {\bibinfo  {journal} {Ann. Rev. Nucl. Part. Sci.}\ }\textbf {\bibinfo
  {volume} {57}},\ \bibinfo {pages} {205} (\bibinfo {year} {2007})},\ \Eprint
  {http://arxiv.org/abs/nucl-ex/0701025} {arXiv:nucl-ex/0701025} \BibitemShut
  {NoStop}%
\bibitem [{\citenamefont {Adler}(2014)}]{PhysRevC.89.044905}%
  \BibitemOpen
  \bibfield  {author} {\bibinfo {author} {\bibfnamefont {e.~a.}\ \bibnamefont
  {Adler}} (\bibinfo {collaboration} {PHENIX Collaboration}),\ }\href {\doibase
  10.1103/PhysRevC.89.044905} {\bibfield  {journal} {\bibinfo  {journal} {Phys.
  Rev. C}\ }\textbf {\bibinfo {volume} {89}},\ \bibinfo {pages} {044905}
  (\bibinfo {year} {2014})}\BibitemShut {NoStop}%
\bibitem [{\citenamefont {Pandit}(2013)}]{FortheSTAR:2013bza}%
  \BibitemOpen
  \bibfield  {author} {\bibinfo {author} {\bibfnamefont {Y.}~\bibnamefont
  {Pandit}} (\bibinfo {collaboration} {the STAR Collaboration}),\ }\href
  {\doibase 10.1088/1742-6596/458/1/012003} {\bibfield  {journal} {\bibinfo
  {journal} {J.Phys.Conf.Ser.}\ }\textbf {\bibinfo {volume} {458}},\ \bibinfo
  {pages} {012003} (\bibinfo {year} {2013})},\ \Eprint
  {http://arxiv.org/abs/1305.0173} {arXiv:1305.0173 [nucl-ex]} \BibitemShut
  {NoStop}%
\bibitem [{\citenamefont {Wang}\ and\ \citenamefont
  {Sorensen}(2014)}]{Wang:2014qxa}%
  \BibitemOpen
  \bibfield  {author} {\bibinfo {author} {\bibfnamefont {H.}~\bibnamefont
  {Wang}}\ and\ \bibinfo {author} {\bibfnamefont {P.}~\bibnamefont {Sorensen}}
  (\bibinfo {collaboration} {STAR Collaboration}),\ }\href@noop {} {\
  (\bibinfo {year} {2014})},\ \Eprint {http://arxiv.org/abs/1406.7522}
  {arXiv:1406.7522 [nucl-ex]} \BibitemShut {NoStop}%
\bibitem [{\citenamefont {Goldschmidt}\ \emph {et~al.}(2015)\citenamefont
  {Goldschmidt}, \citenamefont {Qiu}, \citenamefont {Shen},\ and\ \citenamefont
  {Heinz}}]{Goldschmidt:2015qya}%
  \BibitemOpen
  \bibfield  {author} {\bibinfo {author} {\bibfnamefont {A.}~\bibnamefont
  {Goldschmidt}}, \bibinfo {author} {\bibfnamefont {Z.}~\bibnamefont {Qiu}},
  \bibinfo {author} {\bibfnamefont {C.}~\bibnamefont {Shen}}, \ and\ \bibinfo
  {author} {\bibfnamefont {U.}~\bibnamefont {Heinz}},\ }\href@noop {} {\
  (\bibinfo {year} {2015})},\ \Eprint {http://arxiv.org/abs/1502.00603}
  {arXiv:1502.00603 [nucl-th]} \BibitemShut {NoStop}%
\bibitem [{\citenamefont {Bzdak}\ \emph {et~al.}(2013)\citenamefont {Bzdak},
  \citenamefont {Schenke}, \citenamefont {Tribedy},\ and\ \citenamefont
  {Venugopalan}}]{Bzdak:2013zma}%
  \BibitemOpen
  \bibfield  {author} {\bibinfo {author} {\bibfnamefont {A.}~\bibnamefont
  {Bzdak}}, \bibinfo {author} {\bibfnamefont {B.}~\bibnamefont {Schenke}},
  \bibinfo {author} {\bibfnamefont {P.}~\bibnamefont {Tribedy}}, \ and\
  \bibinfo {author} {\bibfnamefont {R.}~\bibnamefont {Venugopalan}},\ }\href
  {\doibase 10.1103/PhysRevC.87.064906} {\bibfield  {journal} {\bibinfo
  {journal} {Phys.Rev.}\ }\textbf {\bibinfo {volume} {C87}},\ \bibinfo {pages}
  {064906} (\bibinfo {year} {2013})},\ \Eprint {http://arxiv.org/abs/1304.3403}
  {arXiv:1304.3403 [nucl-th]} \BibitemShut {NoStop}%
\bibitem [{\citenamefont {d'Enterria}\ \emph {et~al.}(2010)\citenamefont
  {d'Enterria}, \citenamefont {Eyyubova}, \citenamefont {Korotkikh},
  \citenamefont {Lokhtin}, \citenamefont {Petrushanko} \emph
  {et~al.}}]{dEnterria:2010hd}%
  \BibitemOpen
  \bibfield  {author} {\bibinfo {author} {\bibfnamefont {D.}~\bibnamefont
  {d'Enterria}}, \bibinfo {author} {\bibfnamefont {G.~K.}\ \bibnamefont
  {Eyyubova}}, \bibinfo {author} {\bibfnamefont {V.}~\bibnamefont {Korotkikh}},
  \bibinfo {author} {\bibfnamefont {I.}~\bibnamefont {Lokhtin}}, \bibinfo
  {author} {\bibfnamefont {S.}~\bibnamefont {Petrushanko}},  \emph {et~al.},\
  }\href {\doibase 10.1140/epjc/s10052-009-1232-7} {\bibfield  {journal}
  {\bibinfo  {journal} {Eur.Phys.J.}\ }\textbf {\bibinfo {volume} {C66}},\
  \bibinfo {pages} {173} (\bibinfo {year} {2010})},\ \Eprint
  {http://arxiv.org/abs/0910.3029} {arXiv:0910.3029 [hep-ph]} \BibitemShut
  {NoStop}%
\bibitem [{\citenamefont {Aamodt}\ \emph {et~al.}(2010)\citenamefont {Aamodt}
  \emph {et~al.}}]{Aamodt:2010ft}%
  \BibitemOpen
  \bibfield  {author} {\bibinfo {author} {\bibfnamefont {K.}~\bibnamefont
  {Aamodt}} \emph {et~al.} (\bibinfo {collaboration} {ALICE Collaboration}),\
  }\href {\doibase 10.1140/epjc/s10052-010-1339-x} {\bibfield  {journal}
  {\bibinfo  {journal} {Eur.Phys.J.}\ }\textbf {\bibinfo {volume} {C68}},\
  \bibinfo {pages} {89} (\bibinfo {year} {2010})},\ \Eprint
  {http://arxiv.org/abs/1004.3034} {arXiv:1004.3034 [hep-ex]} \BibitemShut
  {NoStop}%
\bibitem [{\citenamefont {Abelev}\ \emph {et~al.}(2014)\citenamefont {Abelev}
  \emph {et~al.}}]{Abelev:2014mda}%
  \BibitemOpen
  \bibfield  {author} {\bibinfo {author} {\bibfnamefont {B.~B.}\ \bibnamefont
  {Abelev}} \emph {et~al.} (\bibinfo {collaboration} {ALICE Collaboration}),\
  }\href {\doibase 10.1103/PhysRevC.90.054901} {\bibfield  {journal} {\bibinfo
  {journal} {Phys.Rev.}\ }\textbf {\bibinfo {volume} {C90}},\ \bibinfo {pages}
  {054901} (\bibinfo {year} {2014})},\ \Eprint {http://arxiv.org/abs/1406.2474}
  {arXiv:1406.2474 [nucl-ex]} \BibitemShut {NoStop}%
\bibitem [{\citenamefont {Bozek}\ and\ \citenamefont
  {Broniowski}(2013)}]{Bozek:2013uha}%
  \BibitemOpen
  \bibfield  {author} {\bibinfo {author} {\bibfnamefont {P.}~\bibnamefont
  {Bozek}}\ and\ \bibinfo {author} {\bibfnamefont {W.}~\bibnamefont
  {Broniowski}},\ }\href {\doibase 10.1103/PhysRevC.88.014903} {\bibfield
  {journal} {\bibinfo  {journal} {Phys.Rev.}\ }\textbf {\bibinfo {volume}
  {C88}},\ \bibinfo {pages} {014903} (\bibinfo {year} {2013})},\ \Eprint
  {http://arxiv.org/abs/1304.3044} {arXiv:1304.3044 [nucl-th]} \BibitemShut
  {NoStop}%
\bibitem [{\citenamefont {Alvioli}\ \emph {et~al.}(2009)\citenamefont
  {Alvioli}, \citenamefont {Drescher},\ and\ \citenamefont
  {Strikman}}]{Alvioli:2009ab}%
  \BibitemOpen
  \bibfield  {author} {\bibinfo {author} {\bibfnamefont {M.}~\bibnamefont
  {Alvioli}}, \bibinfo {author} {\bibfnamefont {H.-J.}\ \bibnamefont
  {Drescher}}, \ and\ \bibinfo {author} {\bibfnamefont {M.}~\bibnamefont
  {Strikman}},\ }\href {\doibase 10.1016/j.physletb.2009.08.067} {\bibfield
  {journal} {\bibinfo  {journal} {Phys.Lett.}\ }\textbf {\bibinfo {volume}
  {B680}},\ \bibinfo {pages} {225} (\bibinfo {year} {2009})},\ \Eprint
  {http://arxiv.org/abs/0905.2670} {arXiv:0905.2670 [nucl-th]} \BibitemShut
  {NoStop}%
\bibitem [{\citenamefont {Bernhard}\ \emph {et~al.}(2015)\citenamefont
  {Bernhard}, \citenamefont {Marcy}, \citenamefont {Coleman-Smith},
  \citenamefont {Huzurbazar}, \citenamefont {Wolpert} \emph
  {et~al.}}]{Bernhard:2015hxa}%
  \BibitemOpen
  \bibfield  {author} {\bibinfo {author} {\bibfnamefont {J.~E.}\ \bibnamefont
  {Bernhard}}, \bibinfo {author} {\bibfnamefont {P.~W.}\ \bibnamefont {Marcy}},
  \bibinfo {author} {\bibfnamefont {C.~E.}\ \bibnamefont {Coleman-Smith}},
  \bibinfo {author} {\bibfnamefont {S.}~\bibnamefont {Huzurbazar}}, \bibinfo
  {author} {\bibfnamefont {R.~L.}\ \bibnamefont {Wolpert}},  \emph {et~al.},\
  }\href@noop {} {\  (\bibinfo {year} {2015})},\ \Eprint
  {http://arxiv.org/abs/1502.00339} {arXiv:1502.00339 [nucl-th]} \BibitemShut
  {NoStop}%
\bibitem [{\citenamefont {Song}\ and\ \citenamefont
  {Heinz}(2008{\natexlab{b}})}]{Song:2008si}%
  \BibitemOpen
  \bibfield  {author} {\bibinfo {author} {\bibfnamefont {H.}~\bibnamefont
  {Song}}\ and\ \bibinfo {author} {\bibfnamefont {U.~W.}\ \bibnamefont
  {Heinz}},\ }\href {\doibase 10.1103/PhysRevC.78.024902} {\bibfield  {journal}
  {\bibinfo  {journal} {Phys.Rev.}\ }\textbf {\bibinfo {volume} {C78}},\
  \bibinfo {pages} {024902} (\bibinfo {year} {2008}{\natexlab{b}})},\ \Eprint
  {http://arxiv.org/abs/0805.1756} {arXiv:0805.1756 [nucl-th]} \BibitemShut
  {NoStop}%
\bibitem [{\citenamefont {Kisiel}\ \emph {et~al.}(2006)\citenamefont {Kisiel},
  \citenamefont {Taluc}, \citenamefont {Broniowski},\ and\ \citenamefont
  {Florkowski}}]{Kisiel:2005hn}%
  \BibitemOpen
  \bibfield  {author} {\bibinfo {author} {\bibfnamefont {A.}~\bibnamefont
  {Kisiel}}, \bibinfo {author} {\bibfnamefont {T.}~\bibnamefont {Taluc}},
  \bibinfo {author} {\bibfnamefont {W.}~\bibnamefont {Broniowski}}, \ and\
  \bibinfo {author} {\bibfnamefont {W.}~\bibnamefont {Florkowski}},\ }\href
  {\doibase 10.1016/j.cpc.2005.11.010} {\bibfield  {journal} {\bibinfo
  {journal} {Comput.Phys.Commun.}\ }\textbf {\bibinfo {volume} {174}},\
  \bibinfo {pages} {669} (\bibinfo {year} {2006})},\ \Eprint
  {http://arxiv.org/abs/nucl-th/0504047} {arXiv:nucl-th/0504047 [nucl-th]}
  \BibitemShut {NoStop}%
\bibitem [{\citenamefont {Chojnacki}\ \emph {et~al.}(2012)\citenamefont
  {Chojnacki}, \citenamefont {Kisiel}, \citenamefont {Florkowski},\ and\
  \citenamefont {Broniowski}}]{Chojnacki:2011hb}%
  \BibitemOpen
  \bibfield  {author} {\bibinfo {author} {\bibfnamefont {M.}~\bibnamefont
  {Chojnacki}}, \bibinfo {author} {\bibfnamefont {A.}~\bibnamefont {Kisiel}},
  \bibinfo {author} {\bibfnamefont {W.}~\bibnamefont {Florkowski}}, \ and\
  \bibinfo {author} {\bibfnamefont {W.}~\bibnamefont {Broniowski}},\ }\href
  {\doibase 10.1016/j.cpc.2011.11.018} {\bibfield  {journal} {\bibinfo
  {journal} {Comput.Phys.Commun.}\ }\textbf {\bibinfo {volume} {183}},\
  \bibinfo {pages} {746} (\bibinfo {year} {2012})},\ \Eprint
  {http://arxiv.org/abs/1102.0273} {arXiv:1102.0273 [nucl-th]} \BibitemShut
  {NoStop}%
\bibitem [{\citenamefont {Voloshin}(2010)}]{Voloshin:2010ut}%
  \BibitemOpen
  \bibfield  {author} {\bibinfo {author} {\bibfnamefont {S.~A.}\ \bibnamefont
  {Voloshin}},\ }\href {\doibase 10.1103/PhysRevLett.105.172301} {\bibfield
  {journal} {\bibinfo  {journal} {Phys.Rev.Lett.}\ }\textbf {\bibinfo {volume}
  {105}},\ \bibinfo {pages} {172301} (\bibinfo {year} {2010})},\ \Eprint
  {http://arxiv.org/abs/1006.1020} {arXiv:1006.1020 [nucl-th]} \BibitemShut
  {NoStop}%
\bibitem [{\citenamefont {Rybczynski}\ \emph {et~al.}(2013)\citenamefont
  {Rybczynski}, \citenamefont {Broniowski},\ and\ \citenamefont
  {Stefanek}}]{Rybczynski:2012av}%
  \BibitemOpen
  \bibfield  {author} {\bibinfo {author} {\bibfnamefont {M.}~\bibnamefont
  {Rybczynski}}, \bibinfo {author} {\bibfnamefont {W.}~\bibnamefont
  {Broniowski}}, \ and\ \bibinfo {author} {\bibfnamefont {G.}~\bibnamefont
  {Stefanek}},\ }\href {\doibase 10.1103/PhysRevC.87.044908} {\bibfield
  {journal} {\bibinfo  {journal} {Phys.Rev.}\ }\textbf {\bibinfo {volume}
  {C87}},\ \bibinfo {pages} {044908} (\bibinfo {year} {2013})},\ \Eprint
  {http://arxiv.org/abs/1211.2537} {arXiv:1211.2537 [nucl-th]} \BibitemShut
  {NoStop}%
\bibitem [{\citenamefont {Schenke}\ \emph
  {et~al.}(2014{\natexlab{b}})\citenamefont {Schenke}, \citenamefont
  {Tribedy},\ and\ \citenamefont {Venugopalan}}]{Schenke:2014tga}%
  \BibitemOpen
  \bibfield  {author} {\bibinfo {author} {\bibfnamefont {B.}~\bibnamefont
  {Schenke}}, \bibinfo {author} {\bibfnamefont {P.}~\bibnamefont {Tribedy}}, \
  and\ \bibinfo {author} {\bibfnamefont {R.}~\bibnamefont {Venugopalan}},\
  }\href {\doibase 10.1103/PhysRevC.89.064908} {\bibfield  {journal} {\bibinfo
  {journal} {Phys.Rev.}\ }\textbf {\bibinfo {volume} {C89}},\ \bibinfo {pages}
  {064908} (\bibinfo {year} {2014}{\natexlab{b}})},\ \Eprint
  {http://arxiv.org/abs/1403.2232} {arXiv:1403.2232 [nucl-th]} \BibitemShut
  {NoStop}%
\end{thebibliography}%

\end{document}